\documentclass{article}
\usepackage[utf8]{inputenc}
\usepackage{color}
\usepackage{amsmath}
\usepackage{accents}
\usepackage{graphicx}
\usepackage{amssymb}  
\usepackage{lineno}
\usepackage{amsmath}
\usepackage{siunitx}
\usepackage{tabularx}
\usepackage{amsmath}
\usepackage{amsthm}
\usepackage{graphicx}
\usepackage{subcaption}
\usepackage{lscape}
\usepackage{verbatim}
\usepackage{color,soul}
\usepackage{xcolor}
\usepackage{tabularx,ragged2e,booktabs,caption,array,multirow,multicol}
\usepackage{csquotes}
\usepackage{mathtools} 
\usepackage{amsthm}  
\usepackage{listings}
\usepackage{amssymb}
\usepackage{latexsym}
\usepackage{epsfig}
\usepackage{float}
\usepackage{xspace} 
\usepackage{float}
\usepackage[english]{babel}
\usepackage[utf8]{inputenc}
\usepackage{setspace}
\usepackage{lineno}
\usepackage[colorlinks,citecolor=blue,urlcolor=blue]{hyperref}
\usepackage{url}
\usepackage[margin=0.9in]{geometry}
\usepackage{authblk}
\usepackage{tcolorbox}

\title{Learning as We Go -- An Examination of the Statistical Accuracy of COVID-19 Daily Death Count Predictions}

\author[a,b]{Roman Marchant}
\author[d]{Noelle I. Samia}
\author[e]{Ori Rosen}
\author[d]{Martin A. Tanner}
\author[a,b,c,*]{Sally Cripps}

\affil[a]{ARC Centre for Data Analytics for Resources and Environments, Australia}
\affil[b]{Centre for Translational Data Science, The University of Sydney, Australia}
\affil[c]{School of Mathematics and Statistics, The University of Sydney, Australia}
\affil[d]{Department of Statistics, Northwestern University, USA}
\affil[e]{Department of Mathematical Sciences,
University of Texas at El Paso, USA}

\affil[*]{Corresponding author: sally.cripps@sydney.edu.au
+61 425-276-967 }
\date{\today}

\begin{document}

\maketitle

\begin{abstract} 
\noindent \doublespacing
\begin{description}
    \item[OBJECTIVE:]  This paper provides a formal evaluation of the predictive performance of a model (and updates) developed by the Institute  for  Health  Metrics  and  Evaluation  (IHME)   for predicting daily deaths  attributed to COVID-19 for the United States.  
    \item[STUDY DESIGN:] To assess the accuracy of the IHME models,  we examine both forecast accuracy, as well as the predictive performance of the 95\% prediction intervals (PI).   
    \item[RESULTS:] The initial model underestimates the uncertainty surrounding the number of daily deaths. Specifically, the true number of next day deaths fell outside the IHME prediction intervals as much as 76\% of the time, in comparison to the expected value of 5\%. 
    Regarding the updated models, our analyses indicate that the April models show little, if any, improvement in the accuracy of the point estimate predictions.  Moreover, while we observe a larger percentage of states having actual values lying inside the 95\% PI's, this observation may be attributed to the widening of the PI's. 
    A major revised model in early May did result in a decrease in the estimated model uncertainty, albeit at the expense of poorer coverage probability. 
    \item[CONCLUSION:]  Our analysis calls into question the usefulness of the predictions to drive policy making and resource allocation.
\end{description}
\end{abstract}
\vspace{1in}


\newpage

\begin{tcolorbox}{\textbf{Highlights}} 
\begin{enumerate}
\item Regarding the initial IHME model, between 51\% and 76\% of states in the USA have actual daily death counts which lie outside the 95\% prediction interval.
\item  The updated IHME models do not show any improvement in the accuracy of point estimate predictions. 
 \item The rather large level of predictive uncertainty implied by the models over the period 4/4 -- 4/29 casts doubt on their usefulness to drive the development of health, social, and economic policies.
\item A major revised model in early May did result in a decrease in the estimated model uncertainty, at the expense of poorer coverage probability, again calling into question the model's reliability as a predictive tool.
\item The discrepancy between the predicted death and the actual death in the USA has serious implications for the USA government’s future planning and provision of ventilators, PPE, and the staffing of medical professionals equipped to respond to this pandemic.
\end{enumerate}
\end{tcolorbox}

\section {Introduction}
A recent model developed at the Institute for Health Metrics and Evaluation (IHME) provides forecasts for ventilator use and hospital beds required for the care of COVID-19 patients on a state-by-state basis throughout the United States over the period March 2020 through August 2020 \cite{Murray2020} See the related website \url{https://covid19.healthdata.org/projections} for interactive data visualizations. In addition, a manuscript and that website provide projections of deaths per day and total deaths throughout this period for the entire US, as well as for the District of Columbia.  The IHME research has received extensive attention in social media, as well as in the mass media \cite{youtube,azad}.    Moreover, this work has influenced policy makers at the highest levels of the United States government, having been mentioned at White House Press conferences, including March 31, 2020 \cite{youtube}. 

 Our goal in this paper is to provide a framework for formally evaluating the predictive validity of the IHME forecasts for COVID-19 outcomes, as data become sequentially available.  We treat the IHME model (and its various updates) as a ``black box'' and examine the projected numbers of deaths per day in light of the ground truth to help quantify the predictive accuracy of the model.  We do not provide a critique of the assumptions made by the IHME model, nor do we suggest any possible modifications to the IHME approach. Moreover, our analysis should not be misconstrued as an investigation of mitigation measures such as social distancing.  We do, however, strongly believe that it is critical to formally document the operating characteristics of the IHME model -- to meet the needs of social and health planners, as well as a baseline of comparison for future models.
 
\section{Methods}
\label{method}
Our  report  examines the quality of the IHME deaths per day predictions for the initial model over the period March 29--April 2,  2020, for a series of irregularly updated IHME models over the period April 4, 2020--April 29, 2020, as well as for a major revision of the model in early May.  For these analyses we use the  actual deaths attributed to COVID-19 as our ground truth -- our source being the number of deaths reported by Johns Hopkins University  \cite{jhudata}.

Each day the IHME model computes a daily prediction and a 95\% posterior interval (PI) for COVID-19 deaths, four months into the future for each state. For example, on March 29 there is a prediction and corresponding PI for March 30 and March 31, while on March 30 there is a prediction and corresponding PI for March 31.  We call the prediction for a day  made on the previous day a ``1-step-ahead'' prediction. Similarly, a prediction for a day made two days in advance is referred to as a ``2-step-ahead'' prediction, while a prediction for a day made $k$ days in advance is called a ``k-step-ahead'' prediction.

To investigate the accuracy of the point predictions, we computed the logit of the absolute value of the percentage error (APE) \cite{brooks}, denoted by LAPE, where the logit of $|x|$ is defined to be $\frac{1}{1+\exp(-|x|)}$. Note that under this logit transformation, LAPE is a normed metric which approaches one for very large percentage discrepancies (especially when the observed count is close to or equal to zero), while equaling 0.5 for those instances with perfectly accurate predictions. Working on the logit scale avoids the need for ad-hoc rules for discarding outliers. When both observed and predicted death counts are equal to zero, the LAPE is equal to 0.5. Boxplots of LAPE values are compared using the Friedman nonparametric test \cite{milty}, which accounts for possible correlation within states over time.

\section{Results}

\subsection{The Initial Model: 3/29--4/2}

\begin{figure}
    \centering
    \begin{subfigure}[b]{0.31\textwidth}
        \includegraphics[width=\textwidth]{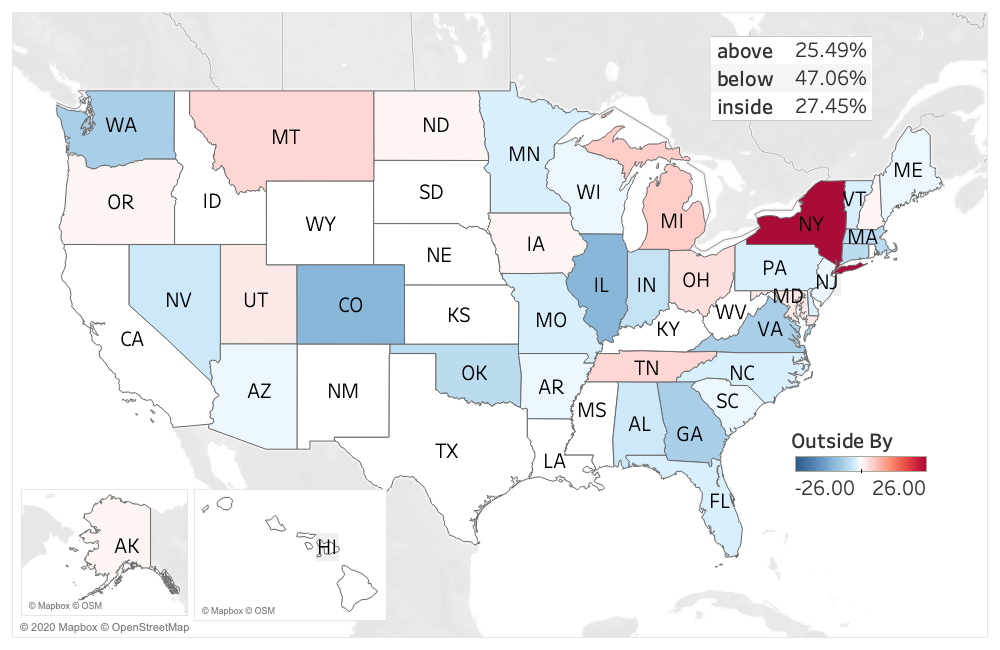}
        \caption{March 30}
        \label{fig_1a}
    \end{subfigure}
    \begin{subfigure}[b]{0.31\textwidth}
        \includegraphics[width=\textwidth]{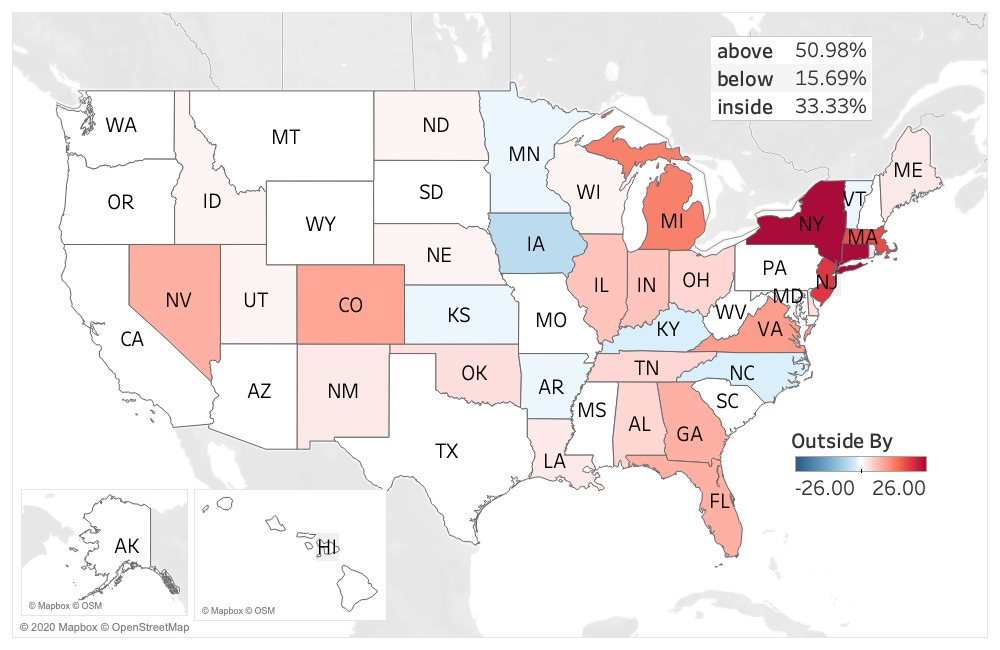}
        \caption{March 31}
        \label{fig_1b}
    \end{subfigure}
    \begin{subfigure}[b]{0.31\textwidth}
        \includegraphics[width=\textwidth]{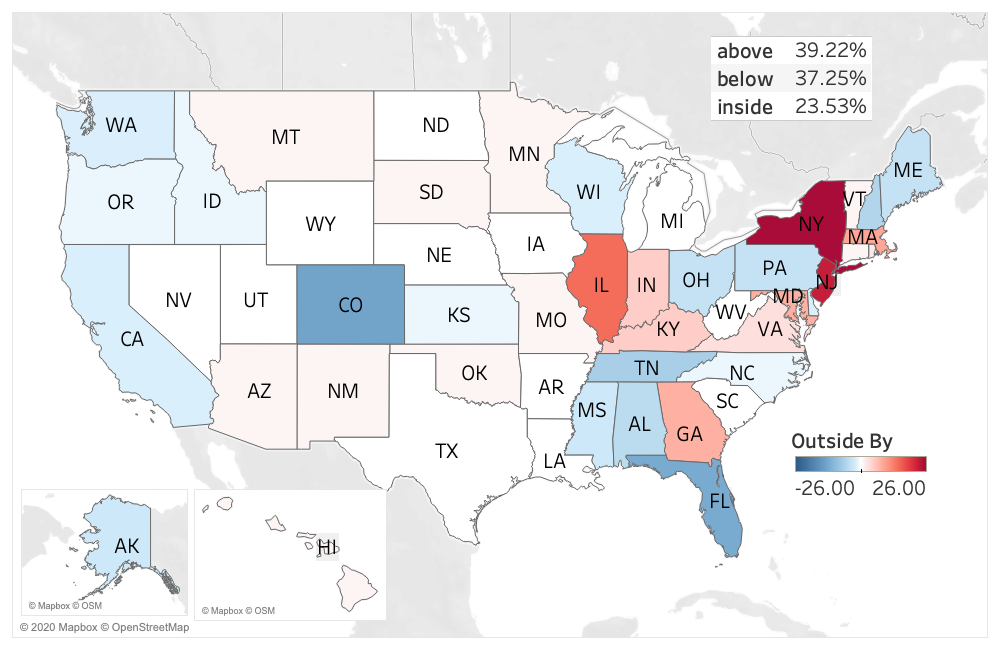}
        \caption{April 1}
        \label{fig_1c}
    \end{subfigure}
    \\
    \begin{subfigure}[b]{0.31\textwidth}
        \includegraphics[width=\textwidth]{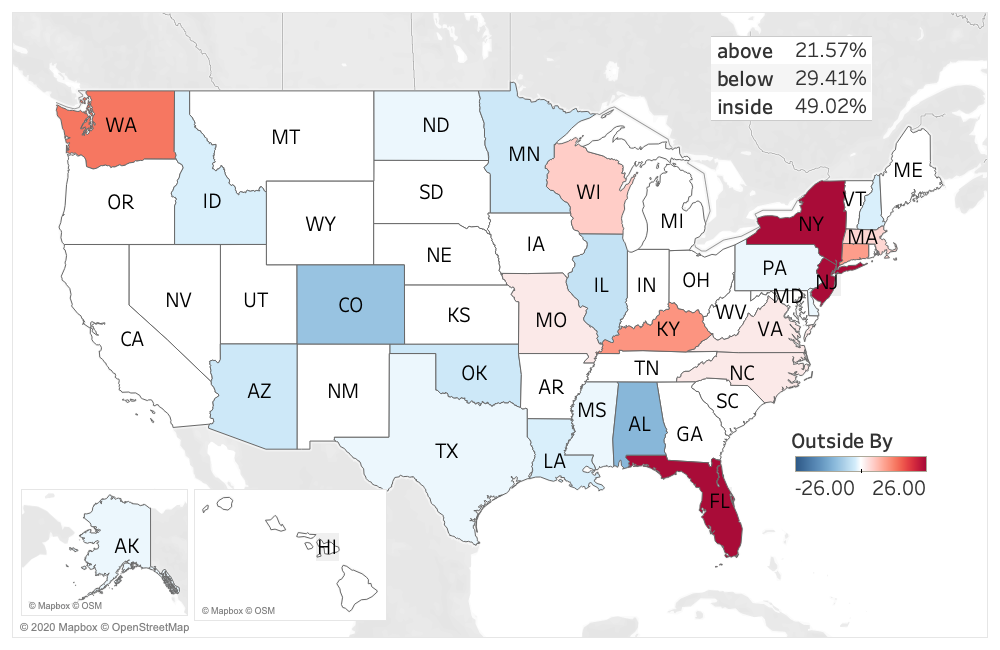}
        \caption{April 2}
        \label{fig_1d}
    \end{subfigure}
        \begin{subfigure}[b]{0.31\textwidth}
        \includegraphics[width=\textwidth]{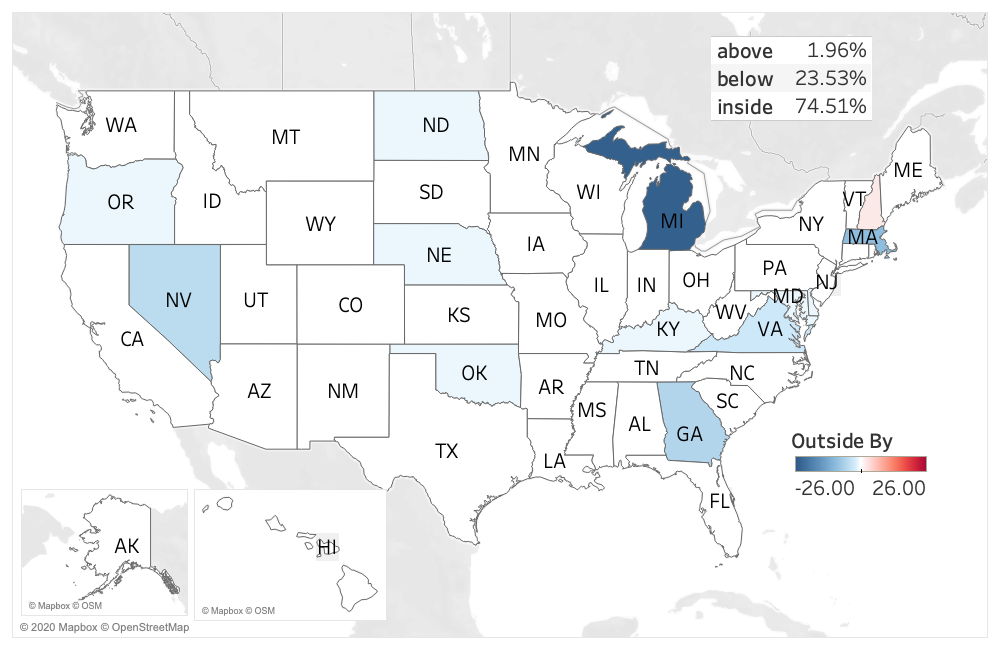}
        \caption{April 4}
        \label{fig_1e}
    \end{subfigure}
        \begin{subfigure}[b]{0.31\textwidth}
        \includegraphics[width=\textwidth]{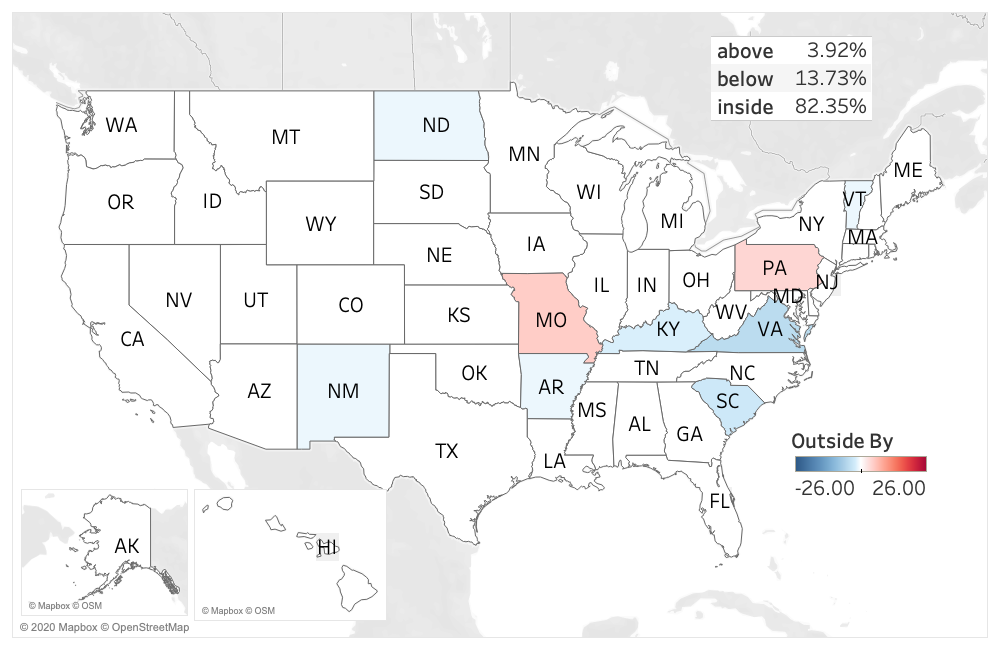}
        \caption{April 8}
        \label{fig_1f}
    \end{subfigure}
    \\
    \begin{subfigure}[b]{0.31\textwidth}
        \includegraphics[width=\textwidth]{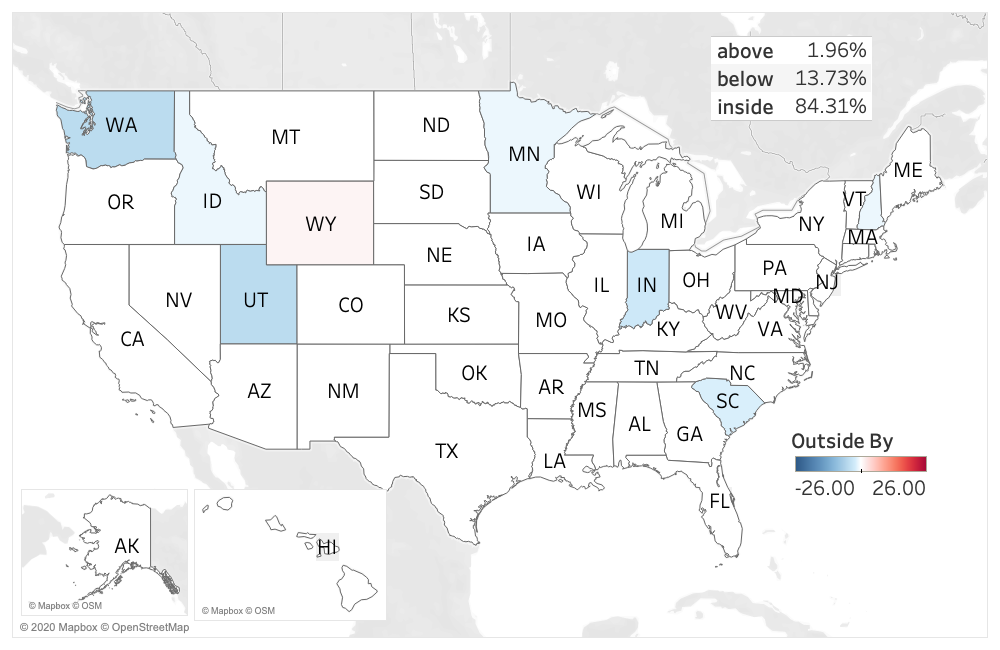}
        \caption{April 13}
        \label{fig_1g}
    \end{subfigure}
        \begin{subfigure}[b]{0.31\textwidth}
        \includegraphics[width=\textwidth]{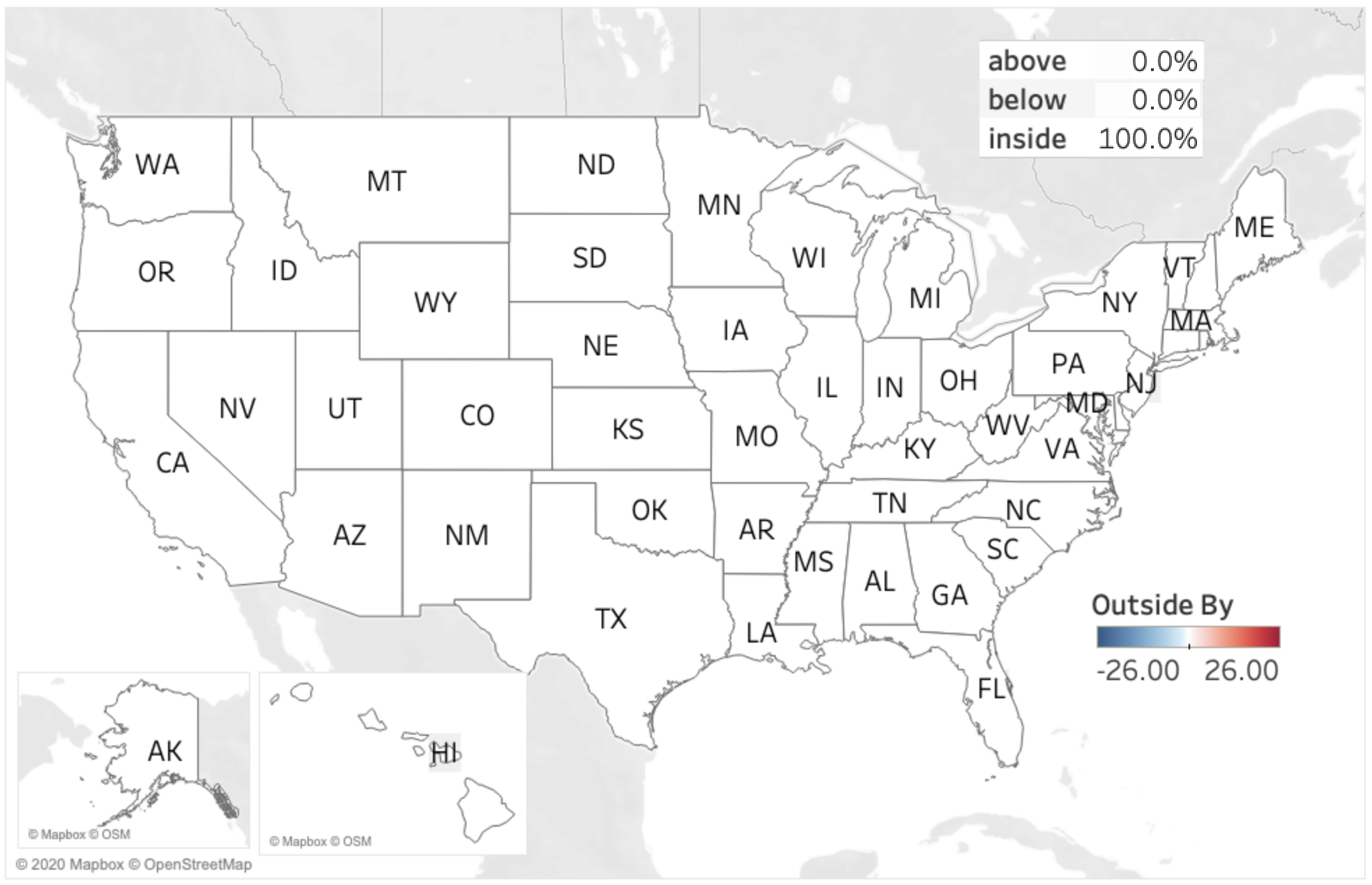}
        \caption{April 17}
        \label{fig_1h}
    \end{subfigure}
        \begin{subfigure}[b]{0.31\textwidth}
        \includegraphics[width=\textwidth]{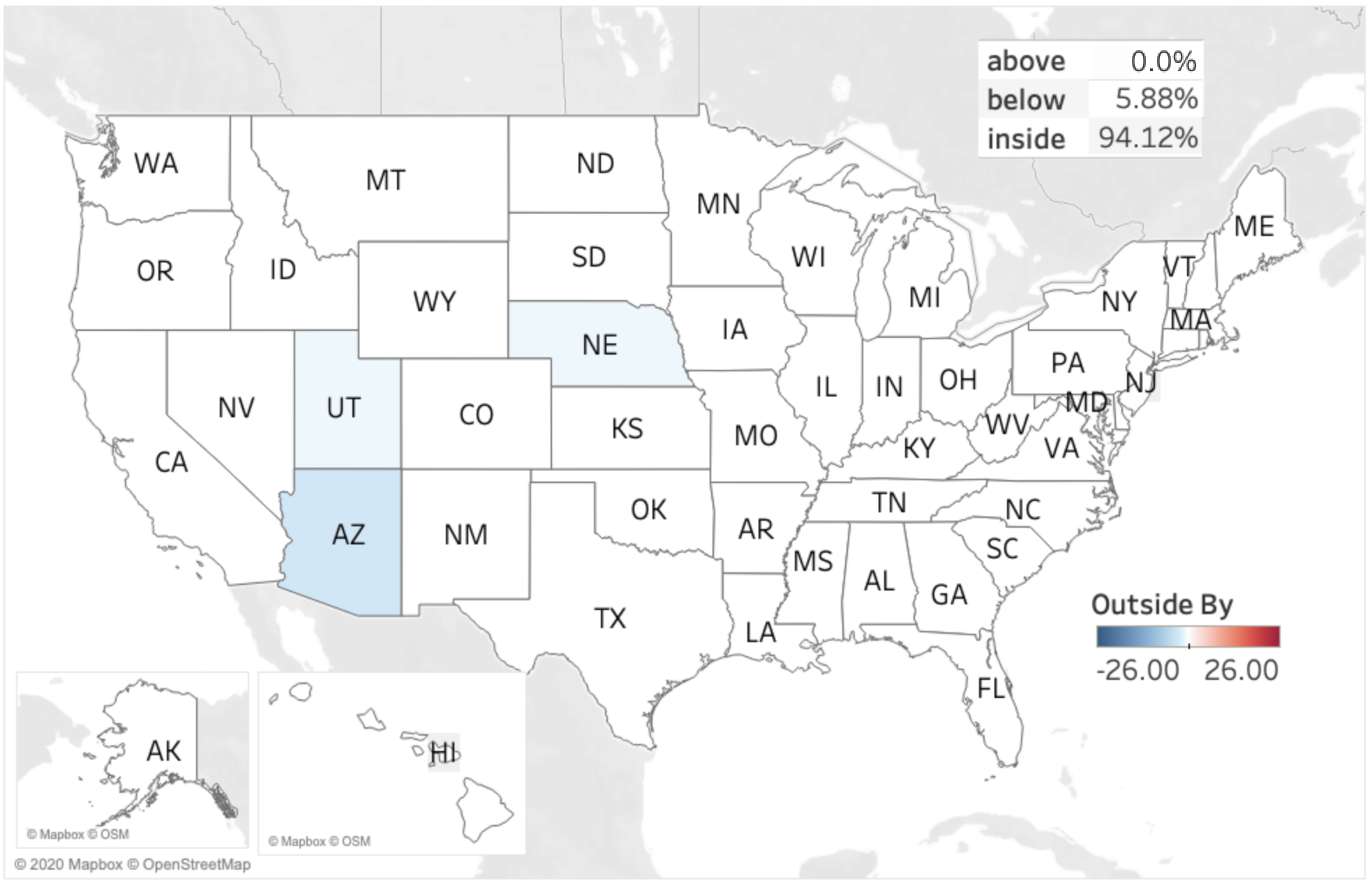}
        \caption{April 28}
        \label{fig_1i}
    \end{subfigure}
    \\
    \begin{subfigure}[b]{0.31\textwidth}
        \includegraphics[width=\textwidth]{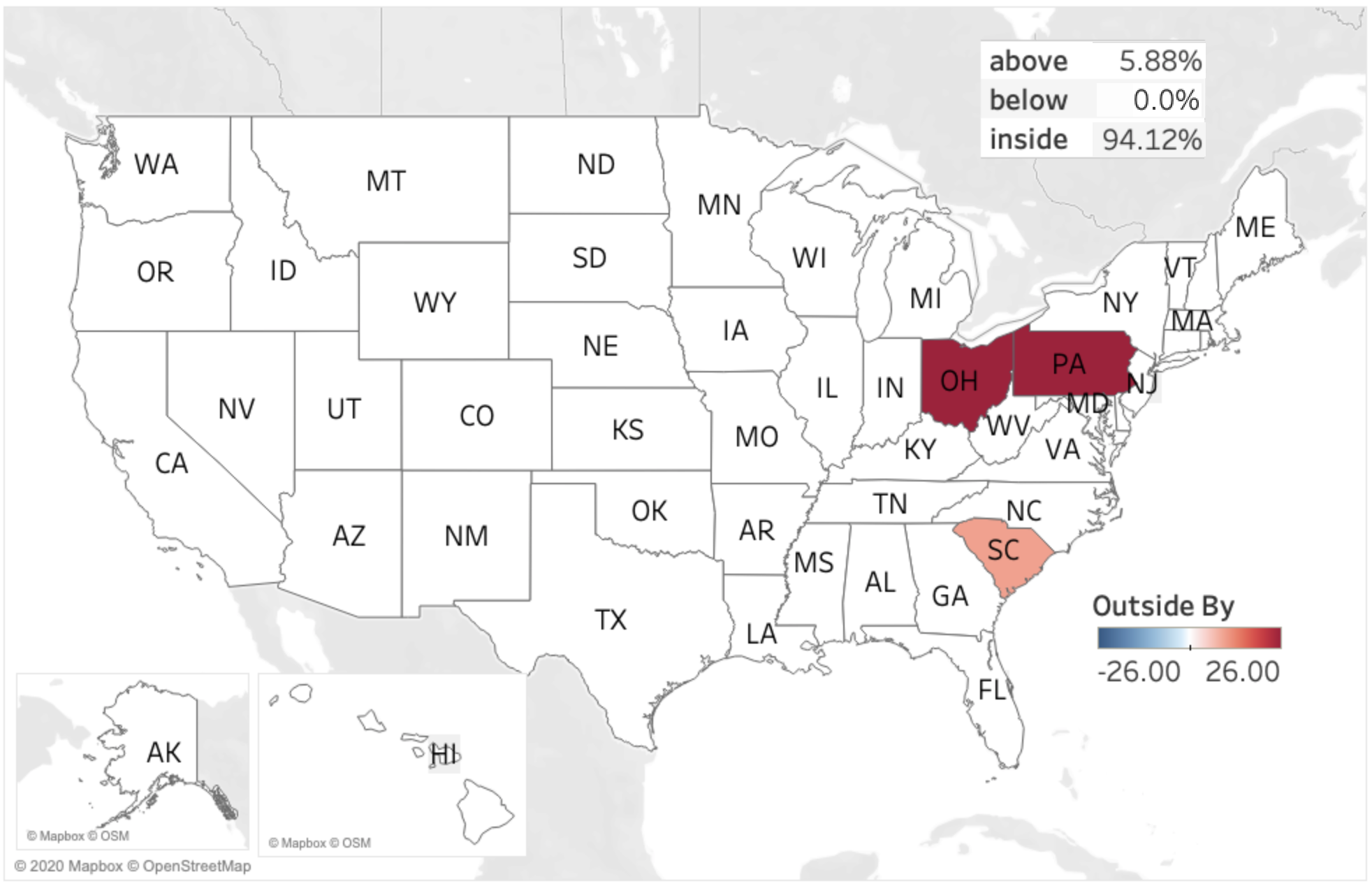}
        \caption{April 29}
        \label{fig_1j}
    \end{subfigure}
        \begin{subfigure}[b]{0.31\textwidth}
        \includegraphics[width=\textwidth]{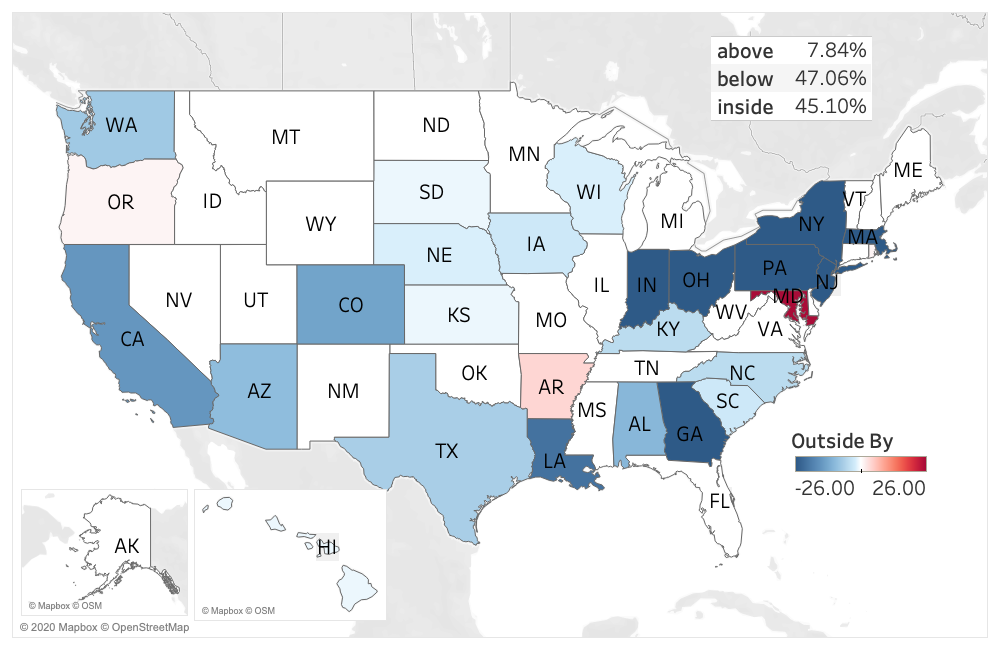}
        \caption{May 2}
        \label{fig_1k}
    \end{subfigure}
    \caption{Discrepancy between actual death counts and one-step-ahead PIs for specific dates (see sub-titles). The color shows whether the actual death counts were less than the lower limit of the 95\% PI (blue), within the 95\% PI (white), or above the upper limit of the 95\% PI (red). The depth of the red/blue color denotes how many actual deaths were above/below the 95\% PI. Initial model: March 30 -- April 2; revised models: April 4 -- April 29; major model update: May 2.}
    \label{fig:maps_1_step_lookahead}
\end{figure}

 Figure~\ref{fig:maps_1_step_lookahead} graphically represents the discrepancy between the actual number of deaths and the 95\% PIs for deaths, by state for the dates March 30 through May 2. The color in these figures shows whether the actual death counts for a state were less than the lower limit of the 95\% PI (blue), or within the 95\% PI (white), or above the upper limit of the 95\% PI (red). The depth of the red/blue color denotes the number of actual death counts above/below  the PI.  A deep red signifies that the number of deaths in that state was substantially  above the upper limit of the 95\% PI, while a light red indicates that the number of deaths was marginally above the 95\% PI upper limit.  Similarly, a deep blue signifies that the number of deaths was  substantially below the lower limit of the 95\% PI, while a light blue indicates that the number of deaths was marginally below 95\% PI lower limit.
 Note that the days in the figure are not consecutive due to the fact that these were the only days for which 1-step ahead predictions were made available by IHME. 
 
 An examination of Figure~\ref{fig:maps_1_step_lookahead} (a) shows that for March 30 only about 27\% of states had an actual number of deaths lying in the 95\% PI for the 1-step-ahead forecast. The corresponding percentages for  March 31, April 1 and April 2, are 33\%, 24\% and 49\%, respectively (see Figures~\ref{fig:maps_1_step_lookahead} (b) -- (d)).  Therefore the percentage of states with actual number of deaths lying outside this interval is 73\%, 67\%, 76\% and 51\% for March 30, March 31, April 1 and April 2, respectively. We note that we would expect only 5\% of observed death counts to lie outside the 95\% PI.
 
For a given day the initial model is also biased, although the direction of the bias is not constant across days.  For the 1-step-ahead prediction for March 30,  47\% of all locations were over-predicted, that is 47\% of all locations had a death count which was below the 95\% PI lower limit, while 26\% were under-predicted. For March 31 the reverse was true; only 16\% of locations had actual death counts below the 95\% PI lower limit while 51\% had actual death counts above the 95\% PI upper limit. This can be clearly seen from Figures 1(a) and 1(b) which are predominantly blue, and red, respectively. See also  the first four lines of Table~\ref{table_errors}.

\begin{table}
\centering
    \begin{tabular}
{ | l | c | c | c | c |}
\hline
Forecast Date & 1 - step & 2 - step & 3 - step & 4 - step \\
\hline
March 30 & 27(47,26) &  &  &  \\
March 31 & 33(16,51) & 45(12,43) &  &  \\
April 01 & 24(37,39) & 37(29,33) & 43(31,25) &  \\
April 02 & 49(29,22) & 50(25,25) & 44(25,31) & 46(25,29) \\
April 03 &  & 31(45,24) & 31(45,24) & 36(39,25) \\
April 04 & 74(24,2) &  & 41(41,18) & 39(43,18) \\
April 05 &  & 84(12,4) &  & 33(51,16) \\
April 06 &  &  & 86(10,4) &  \\
April 07 &  &  &  & 92(4,4) \\
April 08 & 82(14,4) &  &  &  \\
April 09 &  & 84(10,6) &  &  \\
April 10 &  &  & 84(12,4) &  \\
April 11 &  &  &  & 94(6,0) \\

April 13 & 84(14,2) &  &  &  \\
April 14 &  & 96(4,0) &  &  \\
April 15 &  &  & 98(2,0) &  \\
April 16 &  &  &  & 94(2,4) \\
April 17 & 100(0,0) &  &  &  \\
April 18 &  & 96(4,0) &  &  \\
April 19 &  &  & 98(2,0) &  \\
April 20 &  &  &  & 94(4,2) \\
April 28 & 94(6,0) &  &  &  \\
April 29 & 94(0,6) & 90(4,6) &  &  \\
April 30 &  & 86(2,12) & 86(2,12) &  \\
May 01 &  &  & 86(4,10) & 86(4,10) \\
May 02 & 45(47,8) &  &  & 88(6,6) \\
May 03 &  & 33(59,8) &  &  \\
May 04 &  &  & 23(71,6) &  \\
May 05 &  &  &  & 55(25,20) \\
\hline
\end{tabular}
    \caption{Percentage of locations with actual death counts inside the 95\% PI,   for the number of forecast periods. The values in parentheses indicate the percentage of locations that were (below,above) the limits of the 95\% PI. Initial model: March 30 -- April 2; revised models: April 4 -- April 29; major model update: May 2.}
    \label{table_errors}
\end{table}

\begin{figure}
    \centering
    \includegraphics[scale=0.5]{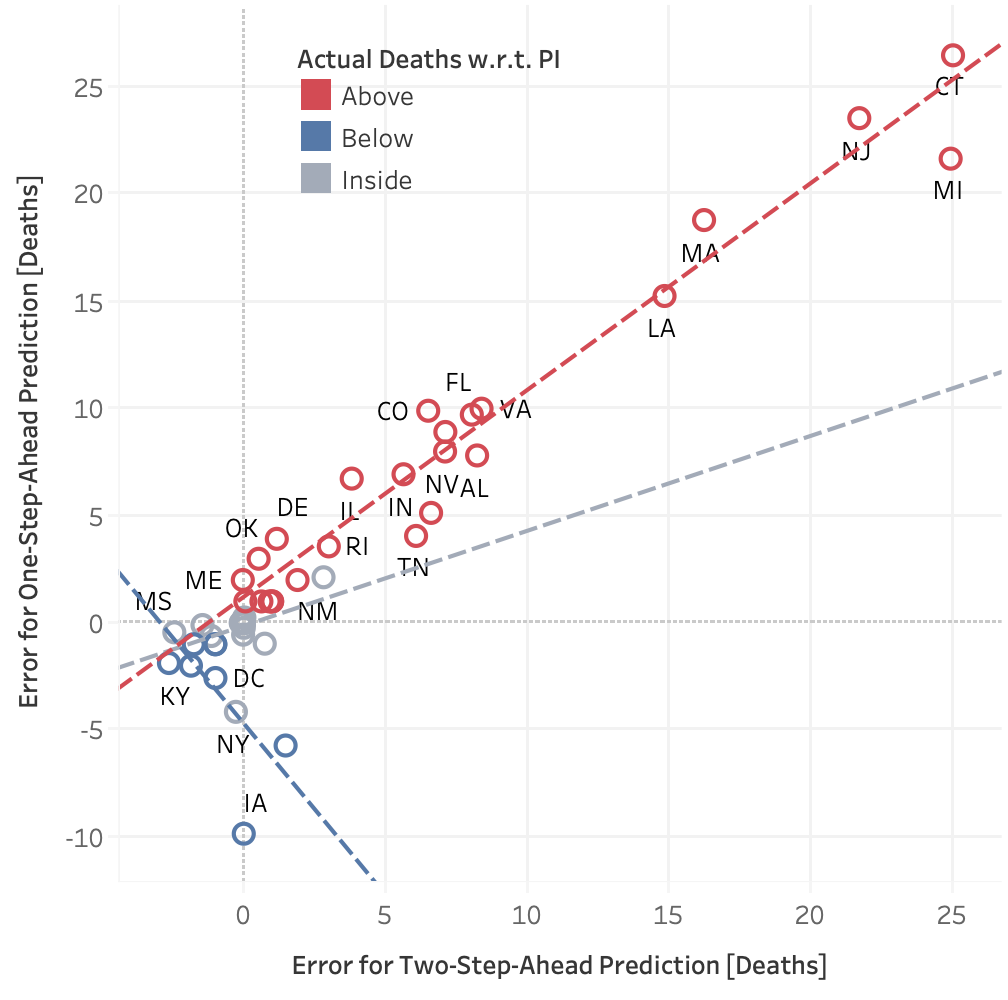}
    \caption{Actual minus  1-step ahead prediction values for March 31, ($y$-axis) vs. actual minus 2-step ahead prediction values for March 31 ($x$-axis).   The colors in the graph correspond to different subsets of the data; red corresponds to those locations where the actual number of deaths was above the 1-step-ahead 95\% PI upper limit, blue corresponds to those locations where the actual number of deaths was below the  1-step-ahead 95\% PI lower limit, while gray corresponds to those locations where the actual number of deaths was within  the the 1-step-ahead 95\% PI.  }
    \label{fig_errors}
\end{figure}

The first four lines of Table~\ref{table_errors} also suggest that the accuracy of predictions does not improve as the forecast horizon decreases, as one would expect. For March 31 and April 1 the forecast accuracy, as measured by the percentage of states whose actual death count lies within the 95\% PI, decreases as the forecast horizon decreases. For March 31, the 2-step ahead prediction is better than the 1-step ahead prediction, while for April 1, the 3-step is better than the 2-step, which in turn is better than the 1-step.  However, April 2 shows  that accuracy slightly improves between the 3-step and the 2-step.

To investigate the relationship between the 2-step-ahead and the 1-step-ahead prediction errors by state, Figure~\ref{fig_errors} shows the March 31 1-step-ahead prediction errors, for predictions made on March 30, on the $y$-axis,  versus the March 31 2-step-ahead prediction errors, for predictions made on March 29, on the $x$-axis.  The colors in the graph correspond to different subsets of the data; red corresponds to those locations where the actual number of deaths was above the 1-step-ahead 95\% PI upper limit, blue corresponds to those locations where the actual number of deaths was below the 1-step-ahead 95\% PI lower limit, while gray corresponds to those locations where the actual number of deaths was within  the 1-step-ahead 95\% PI.   This graph shows a very strong linear association between the predicted errors for the red locations ($R^2 = 96\%,$ $n=25$).  This suggests that the additional information contained in the March 30 data did little to improve the prediction for those locations where the actual death count was much higher than the predicted number of deaths.  The number of observations in the other two subsets of data was insufficient to draw any firm conclusions.


\subsection{The Updated Models: 4/4--5/2}
Per the IHME website, the IHME model underwent a series of updates beginning in early April, followed by a ``major update'' (per the IHME website) in early May. In this subsection we examine the performance of these later versions of the model. Our analysis focuses on two aspects of the IHME model predictions, first on the accuracy of the point estimates used for forecasting and second on the estimated uncertainties surrounding those forecasts.

We note that there are two ways in which the accuracy of the model, as measured by the percentage of states with death counts which fall within  the 95\% PI, can improve.  Either the estimated uncertainty increases and therefore the prediction intervals become much wider, or the estimated expected value improves.  The latter is preferable but much harder to achieve in practice. The former can potentially lead to prediction intervals that are too wide to be useful to drive the development of health, social, and economic policies.

\subsubsection{Uncertainty Estimates of the Updated Models: 4/4--5/2}

A major concern with the initial model had to do with the fact that the PI's had poor coverage - namely, as low as 24\% of the 95\% PI's contained the true value.  We now turn to the evaluation of the uncertainty estimates produced by the updated models from 4/4 -- 5/2.

An examination of Figures~\ref{fig:maps_1_step_lookahead} (e) -- (j), corresponding to the revised models of 4/4 -- 4/29, illustrates that many more states now have actual death counts which lie within the 1-step ahead 95\% PI, as estimated by these revised models than as estimated by the initial model. (Though in this regard, it is noted that any inside percentage presented in Figures~\ref{fig:maps_1_step_lookahead} (e) -- (j), as well as lines 6 -- 26 of Table \ref{table_errors}, below 88\% is statistically significantly different from 0.95 at the 5\% level, according to a one-tailed binomial test.)  In this way, we see that the percentage coverage improved substantially for the April models and lines 6 - 26 of Table \ref{table_errors} confirm this. Interestingly, on  April 29, the model grossly underestimated the death counts in the neighboring states of Ohio and Pennsylvania; see Figure~\ref{fig:maps_1_step_lookahead} (j). 

Regarding the major model update of 5/2, we see in Figure~\ref{fig:maps_1_step_lookahead} (k), a serious deterioration in the empirical coverage probability, as was noted with the initial models. For this 5/2 model, we see from lines 26 -- 29 of Table \ref{table_errors} that the model systematically overestimates the daily death total. 

\begin{figure}
    \centering
    \includegraphics[width=.9\textwidth]{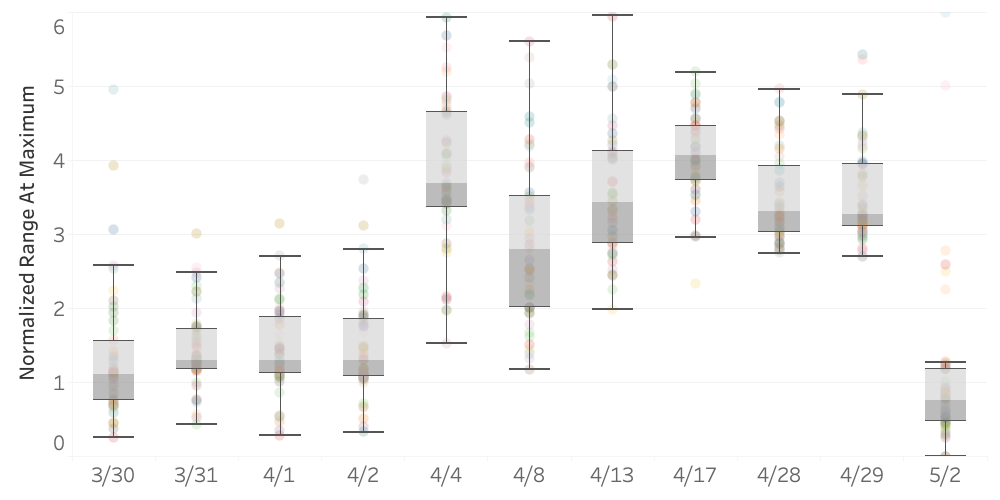}
    \caption{The range at the maximum predicted number of deaths, divided by the maximum predicted number of deaths across states. Each observation represents a state and boxplots are calculated across model release dates. Initial model: March 30 -- April 2; revised models: April 4 -- April 29; major model update: May 2.}
    \label{fig_RANGE}
\end{figure}

To explore this change in the uncertainty estimates of the predictions from the initial model to the revised models to the major model update, we computed the range of the 95\% PI at the date of the forecast peak of daily deaths for each state divided by the predicted value of the number of daily deaths at that peak (analogous to a coefficient of variation). In particular, division by the expected value of daily deaths at the  peak takes into account the fact that those states with higher predicted peak daily deaths will have a larger 95\% PI than those states with a lower expected peak daily deaths.  Figure~\ref{fig_RANGE} presents boxplots of this quantity for all states for both the initial and updated models,   As can be seen from this figure, the normalized range of the PI's expands dramatically with the revised models (4/4--4/29), with $p < 0.001$ according to the Friedman nonparametric test  \cite{milty}.  Apparently, the major model revision of 5/2 resulted in a reduced estimate of variation in the predicted death count.

\subsubsection{Accuracy of Predictions of the Updated Models: 4/4--5/2}
\begin{figure}
    \centering
    \begin{subfigure}[b]{0.31\textwidth}
        \includegraphics[width=\textwidth]{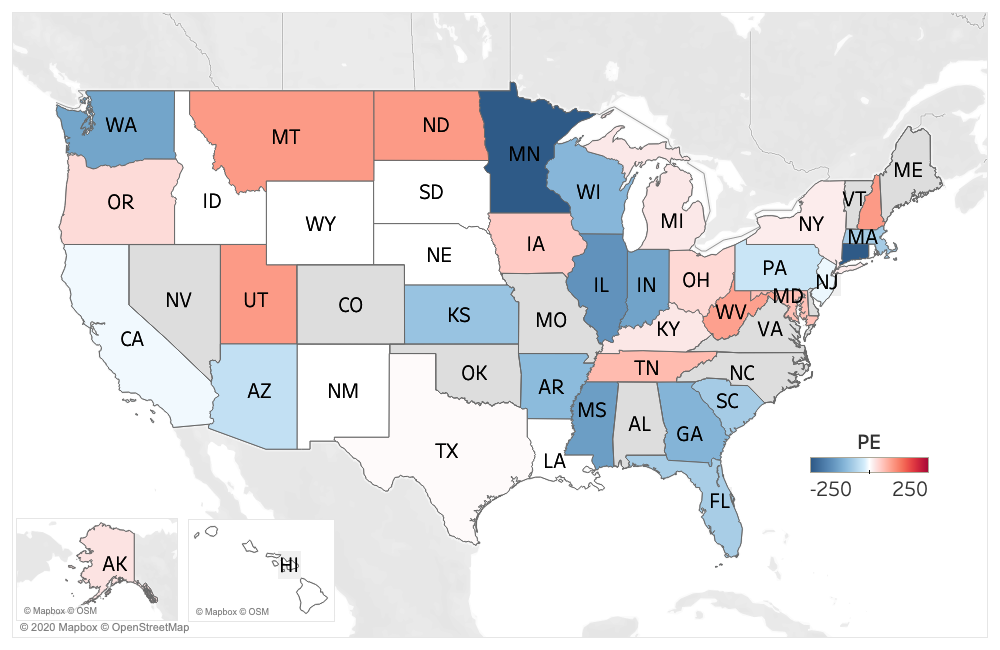}
        \caption{March 30}
        \label{fig:maps_1_step_lookahead_m1_a}
    \end{subfigure}
    \begin{subfigure}[b]{0.31\textwidth}
        \includegraphics[width=\textwidth]{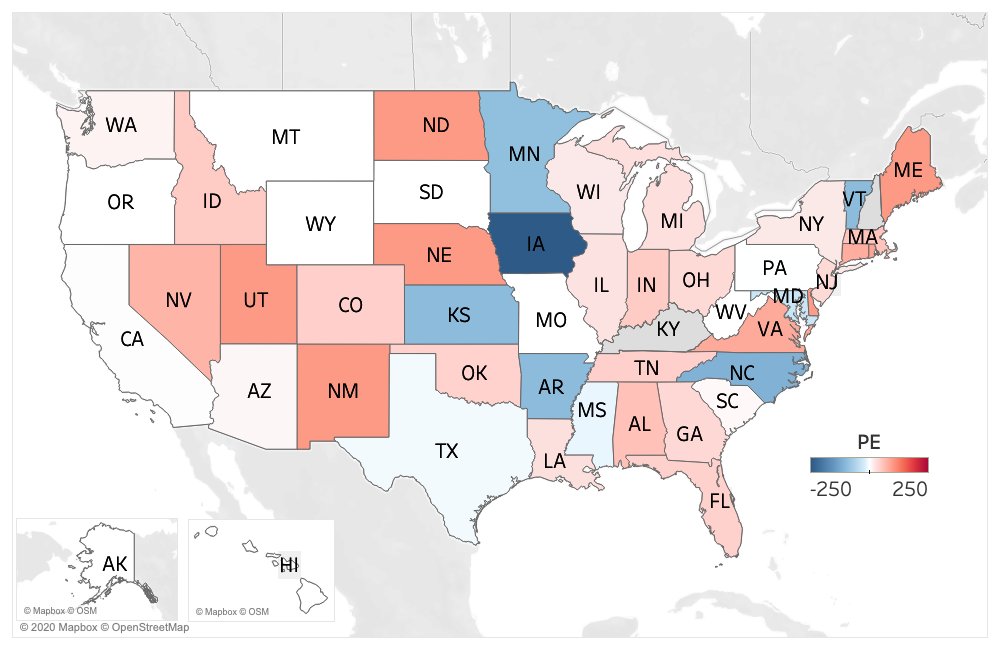}
        \caption{March 31}
        \label{fig:maps_1_step_lookahead_m1_b}
    \end{subfigure}
    \begin{subfigure}[b]{0.31\textwidth}
        \includegraphics[width=\textwidth]{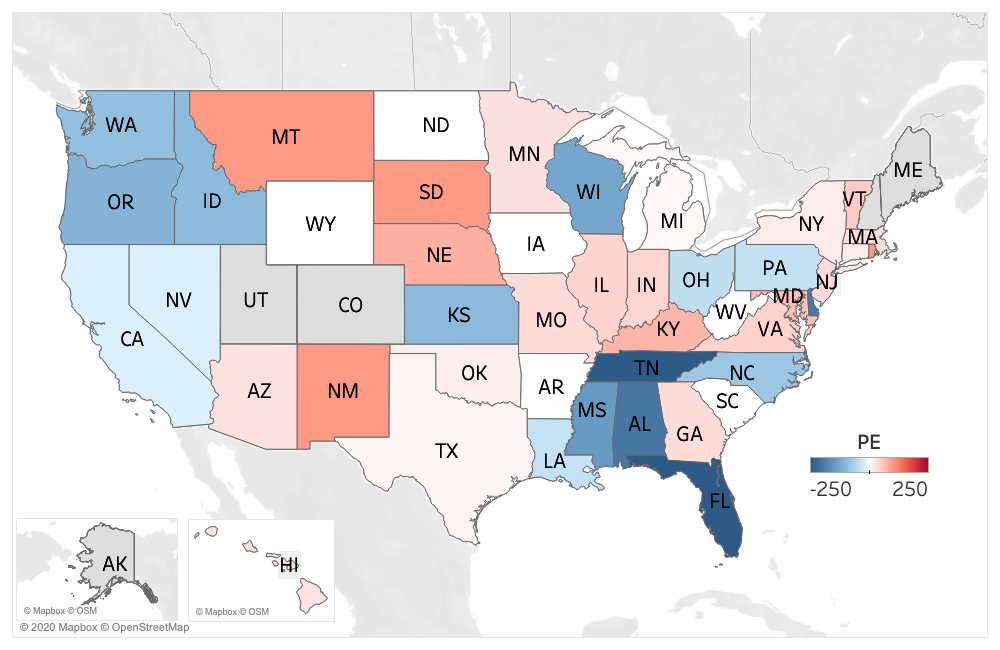}
        \caption{April 1}
        \label{fig:maps_1_step_lookahead_m1_c}
    \end{subfigure}
    \\
    \begin{subfigure}[b]{0.31\textwidth}
        \includegraphics[width=\textwidth]{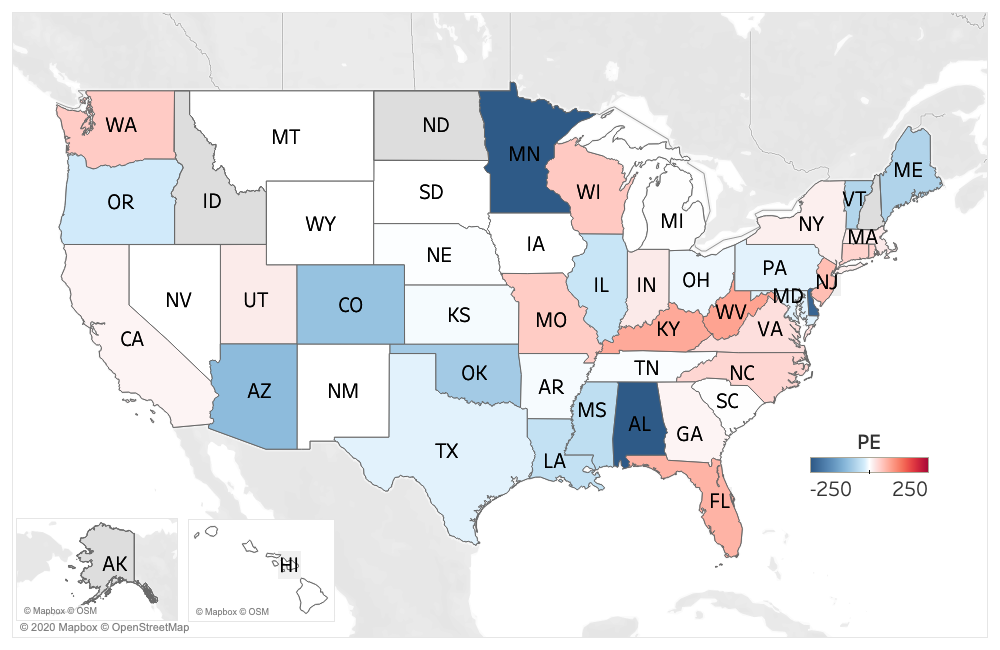}
        \caption{April 2}
        \label{fig:maps_1_step_lookahead_m1_d}
    \end{subfigure}
    \begin{subfigure}[b]{0.31\textwidth}
        \includegraphics[width=\textwidth]{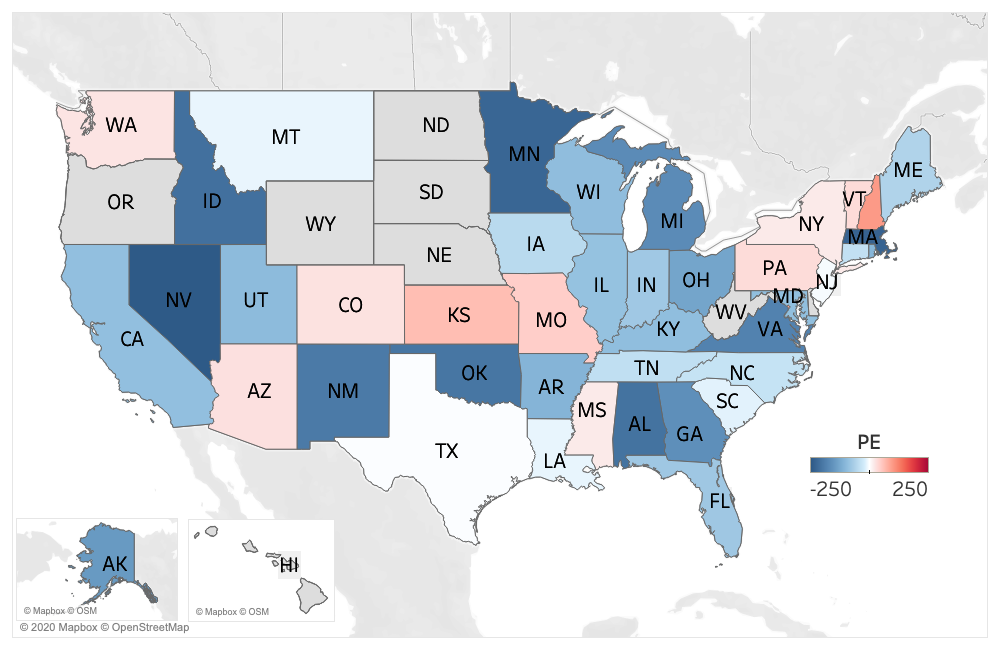}
        \caption{April 4}
        \label{fig:maps_1_step_lookahead_m1_e}
    \end{subfigure}
    \begin{subfigure}[b]{0.31\textwidth}
        \includegraphics[width=\textwidth]{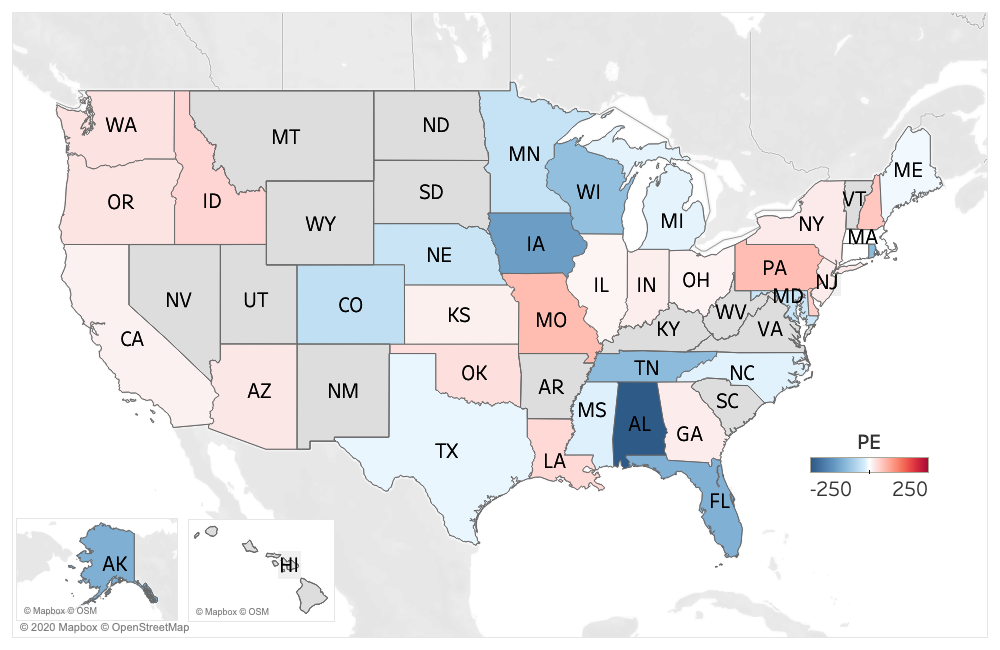}
        \caption{April 8}
        \label{fig:maps_1_step_lookahead_m1_f}
    \end{subfigure}
    \\
    \begin{subfigure}[b]{0.31\textwidth}
        \includegraphics[width=\textwidth]{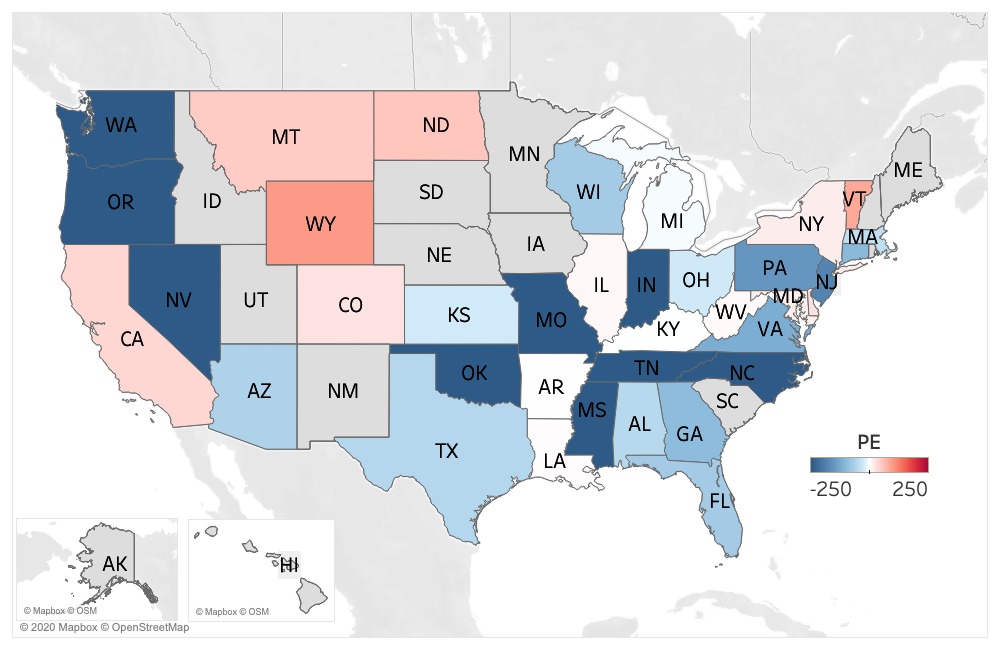}
        \caption{April 13}
        \label{fig:maps_1_step_lookahead_m1_g}
    \end{subfigure}
    \begin{subfigure}[b]{0.31\textwidth}
        \includegraphics[width=\textwidth]{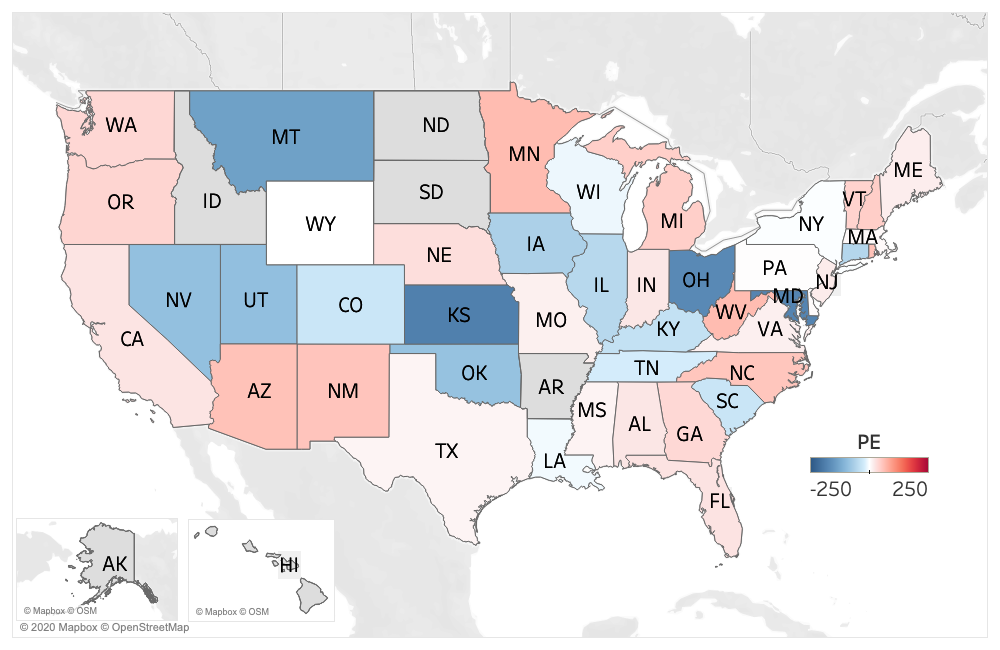}
        \caption{April 17}
        \label{fig:maps_1_step_lookahead_m1_h}
    \end{subfigure}
        \begin{subfigure}[b]{0.31\textwidth}
        \includegraphics[width=\textwidth]{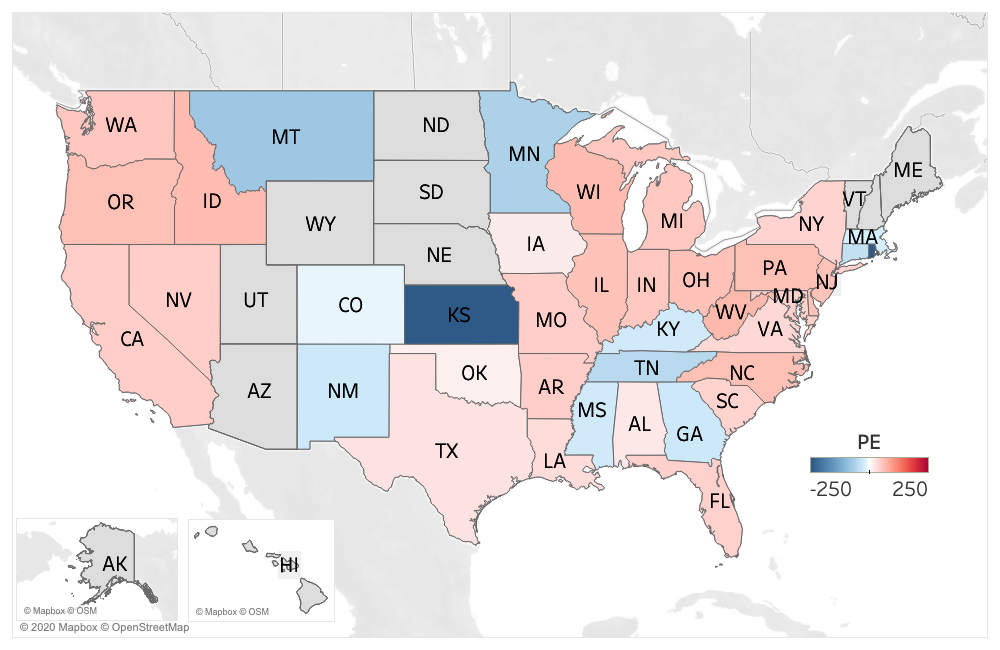}
        \caption{April 28}
        \label{fig:maps_1_step_lookahead_m1_i}
    \end{subfigure}
    \\
        \begin{subfigure}[b]{0.31\textwidth}
        \includegraphics[width=\textwidth]{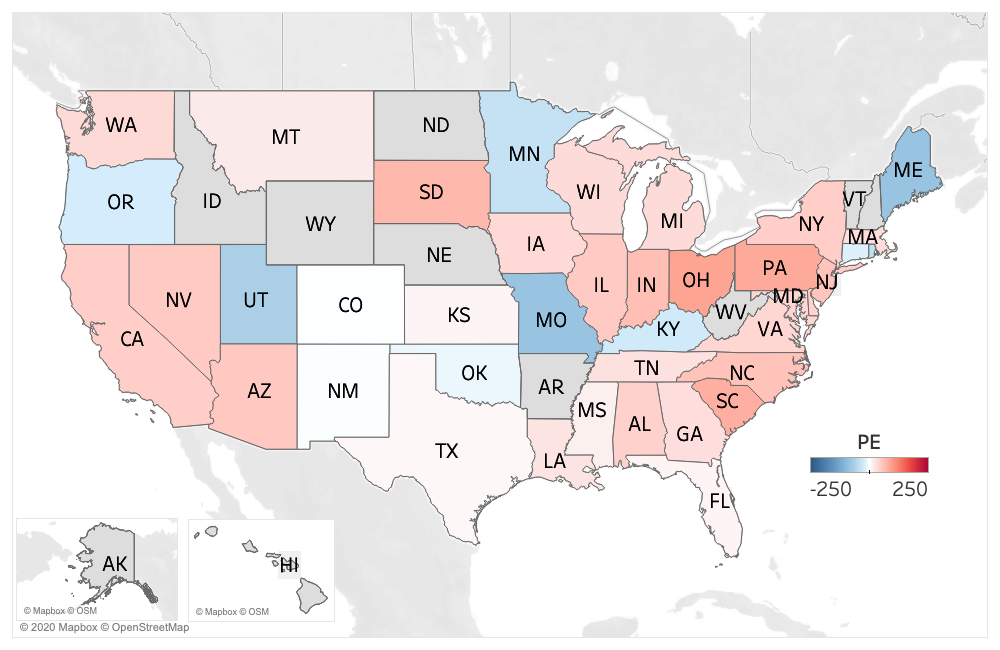}
        \caption{April 29}
        \label{fig:maps_1_step_lookahead_m1_j}
    \end{subfigure}
        \begin{subfigure}[b]{0.31\textwidth}
        \includegraphics[width=\textwidth]{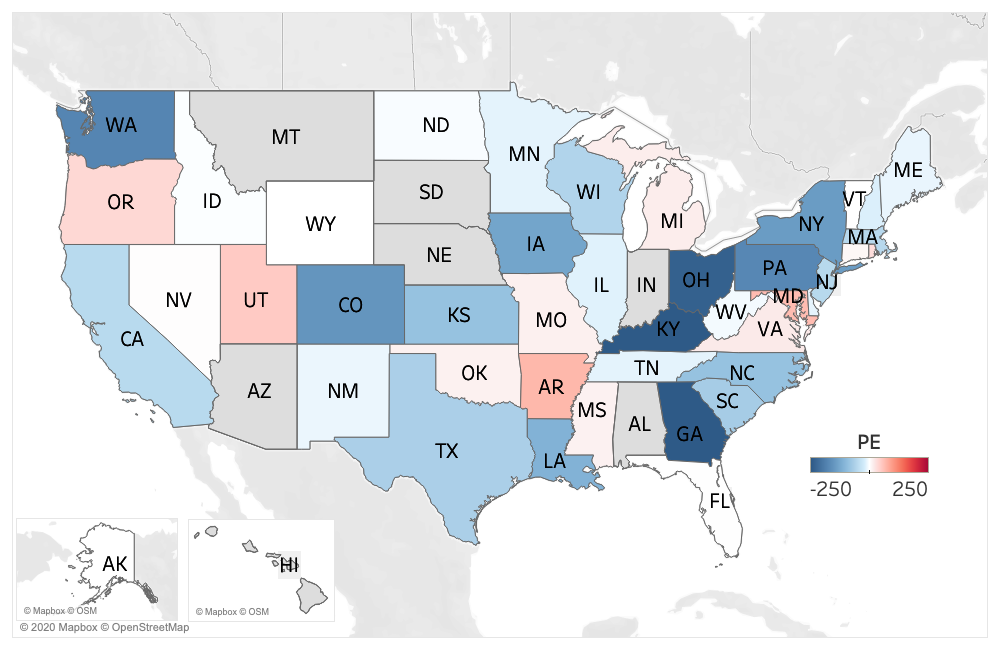}
        \caption{May 2}
        \label{fig:maps_1_step_lookahead_m1_k}
    \end{subfigure}
    \caption{Heat maps of the percentage error (PE) between the actual daily death count and the 1-step ahead predicted daily death count produced by the model for each state, expressed as a percentage of the actual daily death count for the days between March 30 and May 2. The colors in this figure are consistent with Figure 1: blue indicating that the actual death counts were less than the predicted 1-step ahead death count and red indicating that the actual death counts were above the predicted 1-step ahead death count. If the actual value for a state was zero but the predicted value value was not, we have labeled ``NA'' for that state and shaded it as gray.  Initial model: March 30 -- April 2; revised models: April 4 -- April 29; major model update: May 2.}
     \label{fig:maps_1_step_lookahead_m1}
\end{figure}

Figure~\ref{fig:maps_1_step_lookahead_m1} is a heat map of the difference between the actual daily death count and the 1-step ahead predicted daily death count produced by the initial, revised, and major updated models for each state, expressed as a percentage of the actual daily death count for the days between March 30--May 2. For future reference, we denote this percentage error as PE.
(Again, note that the days in the figure are not consecutive due to the fact that these were the only days for which 1-step ahead predictions were made available by IHME.) This graph reproduces Figure~\ref{fig:maps_1_step_lookahead} with two changes. First, instead of analyzing the discrepancy between actual daily deaths and the predicted daily deaths, we now analyze the discrepancy as a percent of the actual daily death count. This is done so that the discrepancy between observed and predicted counts is normalized across different states and on different days. If the actual value and  the predicted value are both zero, we have set the percentage error to zero. If the actual value for a state was zero but the predicted value value was not, we have labeled ``NA'' for that state and shaded it as gray. The second alteration is that the white color coding of states for which the actual death rate was within the 95\% posterior interval is now omitted, so that  Figure~\ref{fig:maps_1_step_lookahead_m1} is now a heat map of the percentage discrepancy.


An examination of Figure~\ref{fig:maps_1_step_lookahead_m1} reveals several features.  First, the initial model produced predictions that were biased toward under-prediction.  The median 1-step PE was greater than or equal to zero for the first four days. This can be seen by the predominance of red in Figures~\ref{fig:maps_1_step_lookahead_m1} (b) and (c),
particularly for March 31 and April 1. The revised models over the next two weeks (particularly April 4 and April 13) had median PE below zero indicating over-prediction, as can be seen by the predominance of blue in Figure~\ref{fig:maps_1_step_lookahead_m1} (e) and (g).  Beginning on April 17, the median PE was positive for the remainder of the month indicative of under-prediction, as noted by the predominance of red in Figures~\ref{fig:maps_1_step_lookahead_m1} (h) -- (j). Following this sustained period of under-prediction, that is, more people died than predicted, the model underwent a major revision.

\begin{figure}
    \centering
    \includegraphics[width=.9\textwidth]{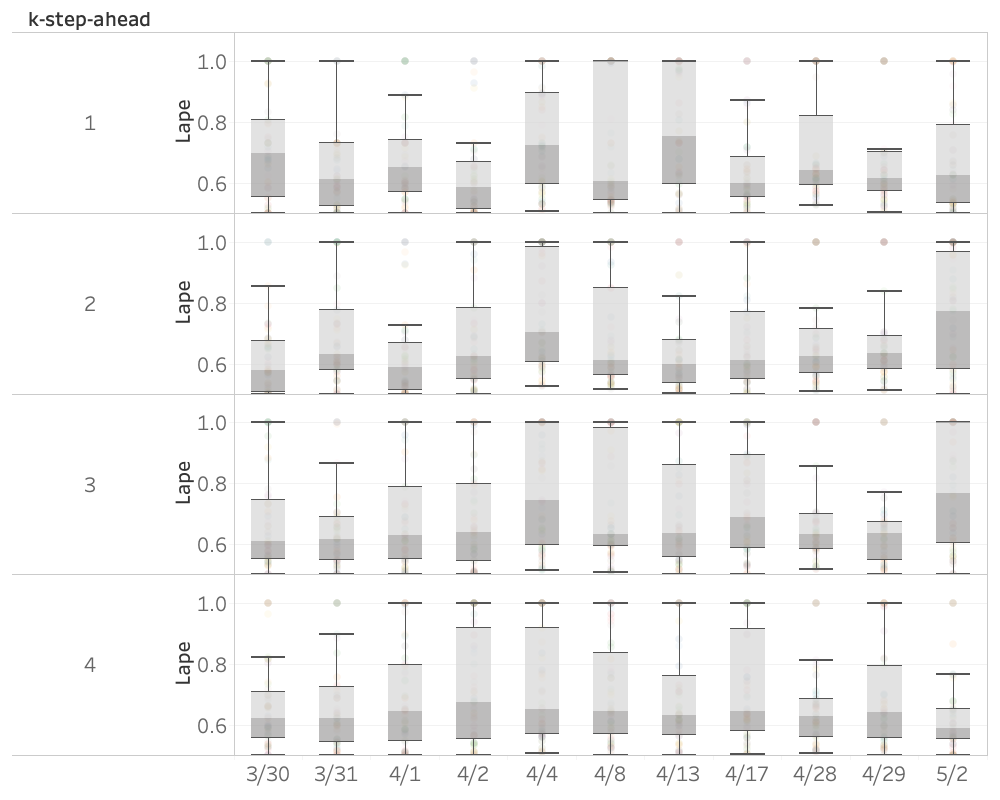}
    \caption{The logit of the absolute percentage error (LAPE) in multiple-step ahead predictions for the model's revision dates. The LAPE values for dates from  April 4 onwards had $k$-step ahead predictions (corresponding to the particular row in the figure) made by the updated models, while those prior to this date had $k$-step ahead predictions made by the initial model. Initial model: March 30 -- April 2; revised models: April 4 -- April 29; major model update: May 2.}
    \label{fig_LAPE}
\end{figure}

Figure~\ref{fig_LAPE} presents boxplots of the LAPE values for the dates March 30 to May 2, corresponding to predictions made with the initial model,  the updated  models, and the major model update of May 2, where each row corresponds to 1-step through 4-step ahead predictions.

An examination of the first row of this figure (corresponding to 1-step ahead predictions) suggests that the predictive performance may have deteriorated somewhat with the updated models, as some boxplots to the right of the initial models seem shifted toward 1.  More formally, the Friedman nonparametric test \cite{milty}, which accounts for possible correlation within states over time, revealed a difference across the eleven time points ($p < 0.001$), with the corresponding post-hoc analysis indicating an elevation in the median LAPE on April 4 and April 13. By the 4-step-ahead prediction (i.e. the fourth row of the figure), the median LAPE values are very similar ($p = 0.40$), where the LAPE values in general take the full range of the logit from 0.5 to 1.0.  As was noted earlier, the prediction accuracy of the initial model on March 30 seems to {\it deteriorate} as the number of steps ahead {i.e. as $k$ \it decreases}.  Interestingly, the major model revision of May 2 seems to follow a similar trajectory.

\section{Discussion and Future Work}

Our results suggest that the initial IHME model substantially underestimated the uncertainty associated with COVID-19 death count predictions. We would expect to see approximately 5\% of the observed number of deaths to fall outside the 95\% prediction intervals. In reality, we found that the observed percentage of death counts that lie outside the 95\% PI to be in the range 49\%--73\%, which is more than  an order of magnitude above the expected percentage. Moreover, we would expect to see 2.5\% of the observed death counts fall above and below the PI. In practice, the observed percentages were asymmetric, with the direction of the bias fluctuating across days.  

In addition, the performance accuracy of the initial model does not improve as the forecast horizon decreases. In fact, Table~\ref{table_errors} indicates that the reverse is generally true. Interestingly, the model's prediction for the state of New York is consistently accurate, while the model's prediction of the neighboring state of New Jersey, which is part of the New York metropolitan area,  is not consistently accurate. 

Our comparison of forecasts made by the initial model versus forecasts to the updated models indicates that the later models do not show any improvement in the accuracy of point predictions.  In fact, there is some evidence that this accuracy has actually decreased.  Moreover, when considering the updated models of early to mid April, while we observe a larger percentage of states having actual values lying inside the 95\% PI, Figure~\ref{fig_RANGE} suggests this observation may be attributed to the widening of the PI's. The width of these intervals does call into question the usefulness of the predictions to drive policy making and resource allocation.  A major model revision in early May resulted in a decrease in the estimated model uncertainty, at the expense of poorer coverage probability.  This observed vacillation  between narrow PI's (low empirical coverage) / wide PI's (high empirical coverage) / narrow PI's (low empirical coverage) reinforces the concern raised by Etzioni \cite{begley}: ``that the IHME model keeps changing is evidence of its lack of reliability as a predictive tool''.   In this regard, see Jewell {\it{et al.}} \cite{Jewell} for general comments as to why the IHME model may suffer from the shortcomings formally documented in the present paper.

In the major update of May, the data reported by IHME was pre-processed to ``smooth'' over fluctuations in the reporting of daily deaths. In particular, the processing involved the following steps: the cumulative number of deaths in a state reported each day's count was replaced with a three-day moving geometric mean.   At this point, the daily deaths were obtained by differencing this processed data.  This idea of pre-processing the data and then fitting a model to the processed data, rather than to the observed death counts, raises several concerns. 

First, the procedure results in replacing each death count by a weighted average of the adjacent death counts, where the weighting extends up to plus/minus 10 days.  Is it reasonable to assume that the difference between the actual data and the processed data is due to reporting errors in all states and across all days of the week?  

Second, there are many methods to pre-process or ``smooth'' data.  Why this method?  Why a window of 3 days?  Why 10 repetitions?    Most importantly, how sensitive are the inferences if another method, or another window size or another number of repetitions were used?  If there is seasonality across the days of the week, why not model this aspect of the data directly, rather than smooth it out?

Third, the prediction intervals of the ``smoothed'' data are akin to confidence intervals for a mean estimate. Indeed the authors make the point they are are interested in the general trend rather than in the variability of the individual daily death count.  However, the correct measure of risk for a decision maker regarding resource allocation on a local level, is the local level of risk, not the risk associated with an average or trend smoothed over potentially twenty days. These concerns (and others) require a comprehensive examination of pre-processing methodology research and are beyond the scope of this paper.  

The accurate quantification of uncertainty in real time is critical for optimal decision making. And while as noted by Jewell {\it{et al.}} \cite{Jewell} ``an appearance of certainty is seductive when the world is desperate to know what lies ahead'', it is perhaps the most pressing issue in policy making that decision makers need accurate assessments of the risks inherent in their decisions. All predictions that are used to inform policy should be accompanied by estimates of uncertainty, and {\it we strongly believe that these estimates should be formally validated against actual data as the data become available -- especially in the case of a novel disease that has affected millions of lives around our entire planet.}

%



\section{Competing Interests} The authors have no competing interests to declare.

\section{Acknowledgements} 
\begin{description}
\item[Authors' contributions:] RM, NIS, OR, MAT and SC contributed to the design of the study, the data analysis and interpretation, and the writing of the paper. 
\item[Funding:] None for the project.
\item[] We thank the authors of the IHME model for making their predictions and data publicly available.  We agree with the statement on their website \url{http://www.healthdata.org/covid/updates}: {\it Having more timely, high-quality data is vital for all modeling endeavors, but its importance is dramatically higher when trying to quantify in real time how a new disease can affect lives.} Without access to the data and predictions this analysis would not have been possible. We also thank Noam B. Tanner for bringing the Brooks reference to our attention. 
\end{description}

\bibliographystyle{plain}
\bibliography{covid19}{}
\end{document}